%% file: letter_accepted_vArXiV.tex
\documentclass[journal,comsoc]{IEEEtran}
\usepackage{amsmath}
\usepackage{amssymb}
\usepackage{graphics}
\usepackage{graphicx}
\usepackage[font=footnotesize]{subcaption}
\usepackage{breqn}
\usepackage{fixltx2e}
\usepackage{epstopdf}
\usepackage[font=footnotesize]{caption}
\usepackage{tikz}
\usetikzlibrary{shapes.geometric,shapes, shapes.misc, patterns, decorations.text,plotmarks,3d}
\usepackage{pgfplots}
\pgfplotsset{compat=newest} 
\pgfplotsset{plot coordinates/math parser=false} 
\newlength\figureheight 
\newlength\figurewidth 
\usepackage{bm}
\usepackage{balance}
\usepackage{cite}
\usepackage{comment}

\makeatletter
\tikzoption{canvas is xy plane at z}[]{%
  \def\tikz@plane@origin{\pgfpointxyz{0}{0}{#1}}%
  \def\tikz@plane@x{\pgfpointxyz{1}{0}{#1}}%
  \def\tikz@plane@y{\pgfpointxyz{0}{1}{#1}}%
  \tikz@canvas@is@plane
}
\makeatother

\tikzset{xzplane/.style={canvas is xz plane at y=#1,very thin}}
\tikzset{yzplane/.style={canvas is yz plane at x=#1,very thin}}
\tikzset{xyplane/.style={canvas is xy plane at z=#1,very thin}}

\ifCLASSINFOpdf
\else
\fi

\newcommand{\m}{\boldsymbol{m}}
\renewcommand{\c}{\boldsymbol{c}}
\newcommand{\e}{\boldsymbol{e}}
\newcommand{\mtilde}{\tilde{\boldsymbol{m}}}
\newcommand{\rr}{\boldsymbol{r}}

\renewcommand{\S}{\mathcal{S}}
\newcommand{\Sd}{\mathcal{S}_{\mathrm{d}}}
\newcommand{\ktilde}{\tilde{k}}

\newcommand{\Dmin}{D_{\rm{min}}}
\newcommand{\GFq}{\text{GF}(q)}
\newcommand{\GFqp}{\text{GF}(q^p)}

\usepackage{amsthm}
\newtheorem{theorem}{Theorem}
\newtheorem{lemma}{Lemma}
\newtheorem{example}{Example}
\newtheorem{definition}{Definition}

\newtheorem{remark}{Remark}

\begin{document}
%
\title{Secure Repairable Fountain Codes}
%
%
%

\author{Siddhartha~Kumar,~\IEEEmembership{Student~Member,~IEEE},
        Eirik~Rosnes,~\IEEEmembership{Senior~Member,~IEEE}, ~and~Alexandre~Graell~i~Amat,~\IEEEmembership{Senior~Member,~IEEE}
\thanks{The work of S.\ Kumar and E.\ Rosnes was partially funded by the Research Council of Norway (grant 240985/F20) and by Simula@UiB. A.\ Graell i Amat was supported by the Swedish Research Council under grant \#2011-5961.}%
\thanks{S.\ Kumar and E.\ Rosnes are with the Department of Informatics, University of Bergen, N-5020 Bergen, Norway, and the Simula Research Lab (e-mail: kumarsi@simula.no; eirik@ii.uib.no).}
\thanks{A.\ Graell i Amat is with the Department of Signals and Systems, Chalmers University of Technology, SE-41296 Gothenburg, Sweden (e-mail: alexandre.graell@chalmers.se).}}


\maketitle

\begin{abstract}

In this letter, we provide the construction of repairable fountain codes (RFCs) for distributed storage systems that are information-theoretically secure against an eavesdropper that has access to the data stored in a subset of the storage nodes and the data downloaded to repair an additional subset of storage nodes. The security is achieved by adding random symbols to the message, which is then encoded by the concatenation of a Gabidulin code and an RFC.
We compare the achievable code rates of the proposed codes with those of secure minimum storage regenerating codes and secure locally repairable codes. 
\end{abstract}


%
\IEEEpeerreviewmaketitle

\section{Introduction}
%
%
%
%


\IEEEPARstart{T}he design of information-theoretically secure distributed storage systems (DSSs)  has attracted a significant interest in the last few years \cite{Sha11, Raw14}. DSSs use erasure correcting codes (ECCs) to yield fault tolerance against storage node failures. The resiliency of the DSS against passive attacks is a good measure of its security. Passive attacks are those where the attacker (referred to as the eavesdropper) gains access to a subset of storage nodes and thereby to partial information on the data stored on the DSS. Information-theoretic security against such attacks involves mixing of information symbols (called the \emph{message}) with random symbols, prior to encoding by an ECC, in a manner such that the eavesdropper does not gain any information about the original message even if he has access to some code symbols. 

Using this idea, \cite{Sha11,Raw14} provided explicit constructions of minimum storage regenerating (MSR) codes that achieve security for an $(\ell_1,\ell_2)$ eavesdropper model where the eavesdropper has access to the content of $\ell_1$ storage nodes and the data that needs to be downloaded to repair $\ell_2$ additional storage nodes. The design of secure locally repairable codes (LRCs) was also addressed in \cite{Raw14}. In particular,  to achieve security, random symbols are appended to the message and the resulting vector of symbols is precoded by a Gabidulin code \cite{Gab85} prior to encoding by an LRC (or MSR code) in \cite{Raw14}. Achieving security comes at the expense of a lower code rate with respect to the original LRC (or MSR code), due to appending random symbols to the message \cite{Raw14}. For the LRC- and MSR-based secure codes, the authors in \cite{Raw14} derived the maximum message size (equivalently, the maximum code rate) that allows to achieve security. Moreover, the code constructions in \cite{Raw14} achieve this maximum. A sufficient condition for the information leakage to the eavesdropper to be zero was also given in \cite{Sha11,Raw14}.

LRCs \cite{Pap12} and MSR codes \cite{Dim10} are appealing code families because they are repair efficient. Repairable fountain codes (RFCs) are another class of repair-efficient ECCs \cite{Ast14}. Like LRCs, they yield a good locality, which implies that few storage nodes are involved in the repair of a failed node. 

In this letter, we present the construction of RFCs that are information-theoretically secure for the $(\ell_1,\ell_2)$ eavesdropper model. As in \cite{Raw14}, we achieve security by appending random symbols to the message and precoding by a Gabidulin code. We prove that the proposed code construction is completely secure for the $(\ell_1,\ell_2)$ eavesdropper model. To prove security, we give a necessary condition for the information leakage to the eavesdropper to be zero, thus extending the sufficient condition in \cite{Sha11,Raw14}. Our proof differs from the one in \cite{Sha11,Raw14} and is based on simple information theory equalities. We compare the achievable code rates (the maximum code rate that allows to achieve security) of the proposed codes with those of secure MSR codes and LRCs in \cite{Raw14}. We show that, for a given rate of the underlying code (RFC, LRC, or MSR code), secure RFCs yield the same achievable code rates as those of secure LRCs and better than those of secure MSR codes when the rate of the underlying code is high enough.

\section{System Model}


We consider a DSS with $n$ storage nodes, each storing one symbol. A message $\m=(m_1,m_2,\ldots,m_k)$, of length $k$ symbols $m_i\in\GFqp$, $i=1,\ldots,k$, where $q$ is a prime and $p$ is a positive integer, is first encoded using an $(n,k)$ ECC of rate $R=k/n$ into a codeword $\bm c=(c_1,c_2,\ldots,c_{n})$ of length $n$. Each of the $n$ code symbols $c_i$, $i=1,\ldots,n$, is then stored into a different storage node. We assume that code symbol $c_i$ is stored in the $i$th storage node and, with a slight abuse of notation, we will refer to both code symbol and storage node $i$ by $c_i$. 
\begin{example}
The bipartite graph shown in Fig.~\ref{Fig: eavesdropper example}(a) represents a message stored on a DSS with $n=6$ storage nodes using a $(6,4)$ ECC. Each code symbol $c_i$, $i=1,\ldots,6$, is a linear combination of its neighboring message symbols $m_i$, $i=1,\ldots,4$ (circles). Each code symbol (squares) is stored on a different storage node. 
\end{example}
\subsection{Security Model}
We consider an $(\ell_1,\ell_2)$ eavesdropper model \cite{Raw14}, where the eavesdropper  can passively observe, but not modify, the content of  $\ell=\ell_1+\ell_2<k$ storage nodes. Out of the $\ell$ nodes, the eavesdropper can observe the symbols stored in a subset of $\ell_1$ storage nodes, which we denote by $\S_1$ ($|\S_1|=\ell_1$). Furthermore, it can observe the data downloaded during the repair of a subset of $\ell_2$ storage nodes, denoted by $\S_2$ ($|\S_2|=\ell_2$), where $\S_1\cap\S_2=\emptyset$. This model is relevant in the scenario where nodes are located at different geographical locations. Peer-to-peer storage systems are examples of such DSSs \cite{Sha11}. We denote the subset of storage nodes from which data is downloaded to repair storage nodes in $\S_2$ by $\Sd$. We will refer to the symbols the eavesdropper obtains as the \emph{eavesdropped symbols}. We also assume that the eavesdropper has perfect knowledge of the ECC used for encoding.

\begin{definition}[\cite{Sha11,Raw14}]
Let $\bm e$ be the vector of eavesdropped symbols that the eavesdropper obtains from the storage nodes in $\S_1\cup\Sd$. A DSS storing a message $\bm m$ (possibly encoded by an ECC) is said to be \emph{completely secure} against an $(\ell_1,\ell_2)$ eavesdropper if the mutual information between the message and the eavesdropped symbols is zero, i.e., $I(\bm m;\bm e)=0$.
\end{definition}

\begin{figure}[t]
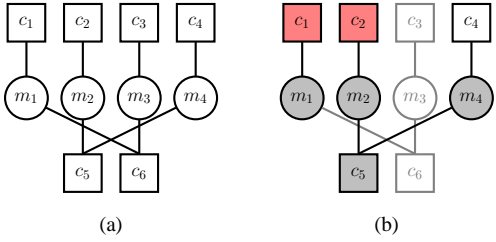

	\centering
	\begin{subfigure}{0.4\columnwidth}
	\centering
	\include{Example_Introduction}
	\vspace{-10pt}
	\caption{}
	\end{subfigure}
	\begin{subfigure}{0.4\columnwidth}
	\centering
	\include{Example_EavesdropperModel}
	\vspace{-10pt}
	\caption{}	
	\end{subfigure}
	\caption{A DSS with 6 storage nodes employing a $(6,4)$ ECC. $\m=(m_1,m_2,m_3,m_4)$ is encoded into the codeword $\c=(c_1,c_2,c_3,c_4,c_5,c_6)$. Each code symbol is stored in a storage node. (a) Bipartite graph of the $(6,4)$ ECC; squares and circles represent code symbols (storage nodes) and message symbols, respectively. (b) Example of a $(1,1)$ eavesdropper, where $\mathcal S_1=\{c_1\}$ and $\mathcal S_2=\{c_2\}$ (in red). Gray symbols are the symbols that the eavesdropper obtains.}
	\label{Fig: eavesdropper example}
	\vspace{-2ex}
\end{figure}

\begin{example}
Fig.~\ref{Fig: eavesdropper example}(b) shows an example of a $(1,1)$ eavesdropper where $\S_1=\{c_1\}$ and $\S_2=\{c_2\}$. Thus, the eavesdropper obtains $c_1=m_1$ and the downloaded data $c_5=m_2+m_4$ and $c_4 = m_4$, and thereby $m_2$, during the repair of $\S_2=\{c_2\}$. In all, the eavesdropper obtains the symbols $m_1,m_2,m_4$, and $c_5=m_2+m_4$, colored in gray in the figure. 
\end{example}

\section{Gabidulin and Repairable Fountain Codes}
We summarize Gabidulin codes and RFCs, which are the building blocks of the secure RFCs presented in Section~\ref{sec:code_construction}.

\subsection{Gabidulin Codes} \label{sec:Gabidulin}
Gabidulin codes are a class of  rank codes \cite{Gab85}. 
An $(N,K)$ Gabidulin code (over GF$(q^p)$) of length $N$, dimension $K$, and minimum rank distance $\Dmin$, can correct up to $\Dmin-1$ rank erasures. 
Gabidulin codes are maximum rank distance codes, i.e., they achieve the Singleton bound, $D_{\rm min}\leq N-K+1$, and are obtained by evaluations of polynomials. More specifically, Gabidulin codes use linearized polynomials. 
\begin{definition} \label{def:fy}
	A linearized polynomial $f(y)$ of degree $t>0$ over GF$(q^p)$ has the form
	\begin{align}
		f(y)=\sum_{i=0}^{t}a_iy^{q^i}, \notag
	\end{align}
	where $a_i\in\text{GF}(q^p)$ and $a_t\neq0$.
	\label{Def: linearPoly}
\end{definition}

A message $\bm m=(m_1, \ldots, m_K)$ is encoded using an $(N,K)$ Gabidulin code as follows.
\begin{enumerate}
	\item Construct a polynomial $f(y)=\sum_{i=1}^{K}m_iy^{q^{i-1}}$.
	\item Evaluate $f(y)$ at $N$  linearly independent (over $\GFq$) points $\{y_1,\ldots, y_N\}\subset \GFqp$ to obtain a codeword $(f(y_1), \ldots, f(y_N))$.
\end{enumerate}
Decoding proceeds as follows.
\begin{enumerate}
	\item Obtain any $K$ evaluations at $K$ linearly independent (over $\GFq$) points. Otherwise, decoding fails.
	\item Perform polynomial interpolation on the $K$ evaluations and recover the original message $\m$ by solving a system of linear equations.
\end{enumerate}

\subsection{Repairable Fountain Codes} \label{sec:RFC}
%

An $(n,k)$ systematic RFC encodes a message $\bm m=(m_1, \ldots, m_k) \in {\rm GF}(q^p)^k$, $q > k$,  into a codeword $\bm c=(c_1,\ldots,c_n)$, where $c_i=m_i$ for $i=1,\ldots,k$. The parity symbols $c_i$, $i=k+1,\ldots,n$, are constructed according to the following three-step procedure.
\begin{enumerate}
	\item Successively select $\xi=O(\log k)$ message symbols independently and uniformly at random with replacement.
	\item For each of the $\xi$ message symbols, a coefficient is drawn uniformly at random from $\GFq \subset {\rm GF}(q^p)$. 
	\item The parity symbol is then obtained as the linear combination of the $\xi$ chosen message symbols, weighted by the corresponding coefficients. 
\end{enumerate}
Each of the $n$ code symbols is stored in a different storage node.
From the code construction, each parity symbol is a weighted sum of at most $\xi$ message symbols. A parity symbol and the corresponding (at most) $\xi$ message symbols is referred to as a local group.
The existence of local groups is a hallmark of any ECC having low locality.
Unlike LRCs, which have only disjoint local groups, RFCs also have overlapping local groups \cite{Ast14}. Furthermore, for each systematic symbol there exist a number of disjoint local groups from which it can be reconstructed. This allows multiple parallel reads of the systematic symbol, accessing the disjoint local groups. When a storage node fails, it is repaired from one of its local groups. This requires the download of at most $\xi$ symbols (from the other at most $\xi$ nodes of the local group).
 Thus, RFCs have low locality, $\xi$, and their repair bandwidth is $\xi p \log q$.
Also, RFCs are near maximum distance separable codes. 
\section{Secure Repairable Fountain Codes}
\label{sec:code_construction}

In this section, we present the construction of RFCs that are secure against the $(\ell_1,\ell_2)$ eavesdropper model. The proposed secure RFCs are obtained by concatenating a Gabidulin code and an RFC. More precisely, consider an $(n,\tilde{k})$ RFC such that each parity symbol is a random linear combination of up to $\xi$ randomly chosen input symbols. Let $\m$ denote the message of length $k=\tilde k-\ell_1-\xi\ell_2$ symbols. A codeword of the proposed secure RFC is constructed as follows.
\begin{enumerate}
	\item Append to $\m$ a random vector $\rr=(r_{1},\ldots,r_{u})$ of length $u=\ell_1+\xi\ell_2$ symbols, drawn independently and uniformly at random from GF$(q^p)$, thus obtaining the vector $\mtilde=(\m,\rr)$.
	\item \emph{Outer code.} Encode $\tilde{\bm m}$ using a $(\tilde k,\tilde k)$ Gabidulin code to obtain the intermediate codeword $\tilde{\bm c}=(\tilde c_1, \ldots, \tilde c_{\tilde{k}})=(f(y_1), \ldots, f(y_{\tilde{k}}))$.
	\item \emph{Inner code.} Encode $\tilde{\bm c}$ using an $(n,\tilde k)$ RFC into the codeword $\bm c=(c_1, \ldots, c_n)$. The $n$ code symbols are then stored in $n$ storage nodes.
\end{enumerate}

\begin{figure}[t]
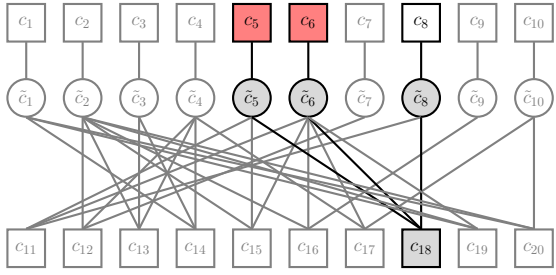

	\centering
	\include{Example_securityV_2}
	\vspace{-2ex}
	\caption{A $(20,10)$ secure RFC. Storage nodes $\mathcal S_1=\{c_6\}$ and $\mathcal S_2=\{c_5\}$ are eavesdropped. Gray symbols are the symbols that the eavesdropper obtains.}
	\vspace{-3ex}
	\label{Fig: Secure repairable fountain code}
\end{figure}

\begin{remark} \label{rem:linear_map}
A $\GFq$-linear combination of evaluations of a linearized polynomial $f(y)=\sum_{i=0}^{t}a_iy^{q^i}$ of some degree $t$ over $\GFqp$ (see Definition~\ref{def:fy}) is itself an evaluation of the same linearized polynomial. In particular,
$\sum_{j=1}^{\kappa} \gamma_j f(\beta_j) = f\left(\sum_{j=1}^{\kappa} \gamma_j \beta_j \right)$,
%
%
where $\kappa$ is a positive integer, $\gamma_j \in \GFq$, and  $\beta_j \in \GFqp$, i.e., $f(\cdot)$ is a linear map over $\GFq$ \cite[Remark~8]{Raw14}. 
Thus, each code symbol $c_i$, $i=1,\ldots,n$, is an evaluation of a linearized polynomial $f(\cdot)$ of degree at most $\tilde{k}-1$ and with coefficients from $\tilde{\bm m}$ at some point $y_i \in \GFqp$, i.e., $c_i = f(y_i)=\sum_{j=1}^{\tilde{k}} \tilde{m}_j y_i^{q^{j-1}}$. 
\end{remark}

\begin{example}
Fig.~\ref{Fig: Secure repairable fountain code} depicts a toy example of a $(20,10)$ secure RFC for a $(1,1)$ eavesdropper. Here, $\mtilde$ comprises $k=\tilde k-\ell_1-\xi\ell_2=6$ message symbols and $u=\ell_1+\xi\ell_2=4$ random symbols. $\mtilde$ is encoded using the concatenation of a $(10,10)$ Gabidulin code and a $(20,10)$ RFC. Due to the outer encoding by the Gabidulin code, each code symbol $c_i$, $i=1,\ldots,20$, is an evaluation of a linearized polynomial (see Remark~\ref{rem:linear_map}). Another consequence is that the final code retains the repair properties of the inner code (the RFC). For this example, the code locality is $\xi=3$. 
\end{example}

In the following, we show that the proposed secure RFCs achieve complete security for the $(\ell_1,\ell_2)$ eavesdropper model. We first prove a sufficient and necessary condition for $I(\m;\e)=0$ using an alternative proof to the one in \cite{Sha11, Raw14}.

\begin{theorem} \label{Lemma: Secrecy lemma}
Let $\m$ be a message which is stored in a DSS by first appending to it a vector $\rr$ of random symbols and then encoding $(\m,\rr)$ by an ECC. Also, let $\bm e$ be the vector of code symbols the eavesdropper has access to. Then, the information leakage to the eavesdropper is zero, i.e., $I(\bm m;\bm e)=0$, if and only if $H(\bm r|\bm e,\bm m)=H(\bm r)- H(\bm e)$.  	

\end{theorem}

\begin{IEEEproof}
	We prove the theorem using simple information theory equalities,
	\begin{align*}
		\begin{split}
			I(\bm m;\bm e)&=H(\bm m)-H(\bm m|\bm e)\\
			&\overset{(a)}{=}H(\bm m)-H(\bm m|\bm e)+H(\bm e|\bm m,\bm r)\\
			&=H(\bm m)-H(\bm m|\bm e)+H(\bm e|\bm m)-I(\bm r; \bm e|\bm m)\\
			&\overset{(b)}{=}H(\bm e)-I(\bm r;\bm e|\bm m)\\
			&=H(\bm e)-H(\bm r|\bm m)+H(\bm r|\bm e, \bm m)\\
			&\overset{(c)}{=}H(\bm e)-H(\bm r)+H(\bm r|\bm e,\bm m),
		\end{split}
	\end{align*}
	where $(a)$ follows from the fact that $H(\bm e|\bm m,\bm r)=0$, since eavesdropped symbols are a function of $\bm m$ and $\bm r$, $(b)$ follows from $H(\bm e)-H(\bm e|\bm m)=H(\bm m)-H(\bm m|\bm e)$, and $(c)$ follows from the fact that $\bm r$ and $\bm m$ are stochastically independent of each other, i.e., $H(\bm r|\bm m)=H(\bm r)$. Thus,
\begin{equation} 
I(\bm m;\bm e)=0 	\Leftrightarrow  H(\bm r|\bm e,\bm m)=H(\bm r)- H(\bm e).
\label{Eq: lemma}
\end{equation}
\end{IEEEproof}

We remark that in \cite{Sha11} and \cite[Lem.~4]{Raw14} a sufficient condition on $I(\m;\e)=0$ was proved, whereas Theorem~\ref{Lemma: Secrecy lemma} gives a sufficient and necessary condition. ECCs for which Theorem~\ref{Lemma: Secrecy lemma} is satisfied do not leak any information, i.e., they are completely secure.  
%
%
%
In Theorem~\ref{th:Security} below, we use the following lemma to prove that our proposed code construction is completely secure for the $(\ell_1,\ell_2)$ eavesdropper model.


\begin{lemma}\label{Lemma: entropy}
Consider the $(\ell_1,\ell_2)$ eavesdropper model. For the proposed code construction (with $u=\ell_1+\xi\ell_2$ random symbols $\bm r$), $H(\bm e)\leq H(\bm r)=(\ell_1+\xi\ell_2)p\log q$, where $\bm e$ is the vector of code symbols the eavesdropper has access to.
\end{lemma}
\begin{IEEEproof}
Consider the repair of a single storage node $c_i$  in $\mathcal S_2$, and let $\Gamma^{(i)}$ denote \emph{the} local group (there are many) used for the repair of storage node $c_i$. Each local group contains one inner code parity symbol and at most $\xi$ inner code message symbols to which it is connected. Thus, $|\Gamma^{(i)}| \leq \xi+1$. Since the inner code parity symbol is a ${\rm GF}(q)$-weighted linear combination of the (at most) $\xi$ inner code message symbols from the local group, $\Gamma^{(i)}$ contains at most $\xi$ stochastically independent symbols. Considering the repair of all storage nodes in $\mathcal S_2$, it follows by the argument above that at most $\xi \ell_2$ stochastically independent inner code symbols are eavesdropped during the repair process. Also, since each storage node stores a single symbol, the eavesdropper has access to an additional $\ell_1$ inner code symbols from the storage nodes in $\mathcal S_1$. Hence, in total, the eavesdropper has access to at most $\ell_1+\xi\ell_2$ stochastically  independent symbols from $\bm c$. 
%
%
%
Thus, $H(\bm e) \leq (\ell_1+\xi\ell_2) p \log q$. 
Furthermore, since $\bm r$ contains $u=\ell_1+\xi\ell_2$ uniform independent random symbols, $H(\bm r)=(\ell_1+\xi\ell_2)p\log q$, and the result follows.
\end{IEEEproof}

\begin{theorem}
\label{th:Security}
The code comprising of a Gabidulin code as its outer code and an RFC as its inner code, which encodes a vector $\tilde{\bm m}=(\bm m,\bm r)$ that consists of a message $\bm m$ of length $k$ and a random vector $\bm r$ of length $u=\ell_1+\xi\ell_2$ is completely secure for the $(\ell_1,\ell_2)$ eavesdropper model.
\end{theorem}

\begin{IEEEproof}
To prove security, we show that $H(\bm r|\bm e,\bm m) = H(\bm r) - H(\bm e)$, which is \emph{equivalent} to $I(\bm m;\bm e)=0$ according to Theorem~\ref{Lemma: Secrecy lemma}. Each eavesdropped symbol $e_i$, $i=1,\dots,w$, where $w = |\bm e|$, corresponds to a code symbol and therefore is an evaluation of $f(\cdot)$ at some point $z_i \in \{y_1,\ldots,y_n\} \subset \GFqp$, where $c_i = f(y_i)$ (see Remark~\ref{rem:linear_map}).  Thus, for $i=1,\ldots,w$,
\begin{equation} 
		e_i = f(z_i)=\sum_{j=1}^{\tilde k}\tilde m_jz_i^{q^{j-1}}=\sum_{j=1}^{k}m_jz_i^{q^{j-1}}+\sum_{j=1}^{u}r_jz_i^{q^{k+j-1}}
		\label{eq:fy}
	\end{equation}
since $\tilde{\bm m} = (\bm m,\bm r)$. In the following, $\bm r = (r_1,\ldots,r_u)$ is assumed to be the unknowns (the message $\bm m$ and the eavesdropper vector $\bm e$ are assumed to be known) in the linear system of equations defined in (\ref{eq:fy}).

Let $1 \leq \nu \leq w$ (by definition) be the number of $\GFq$-linear independent symbols  of $\{e_1,\ldots,e_{w}\}$, denoted by $\tilde{\bm{e}} = (\tilde e_1,\ldots,\tilde e_{\nu})$. The corresponding vector of points from $\{z_1,\ldots,z_{w}\}$ is denoted by $\tilde{\bm{z}} = (\tilde z_1,\ldots,\tilde z_{\nu})$. From (\ref{eq:fy}), $\tilde{\bm e} =  \bm b(\tilde{\bm{z}},\bm m) + \bm{r} \cdot \bm{A}(\tilde{\bm{z}})$, where $\bm b(\tilde{\bm{z}},\bm m)$  is a length-$\nu$ row vector and $\bm{A}(\tilde{\bm{z}})$ is a $u \times \nu$ matrix. Since $\{\tilde e_1,\ldots,\tilde e_{\nu}\}$ are $\GFq$-linear independent, the matrix $\bm{A}(\tilde{\bm{z}})$ is of full column-rank (i.e., its column space is a vector space over $\GFq$ of dimension $\nu$), and since the $u$ random symbols in $\bm r$ are chosen independently and uniformly at random from ${\rm GF}(q^p)$, $\{\tilde e_1,\ldots,\tilde e_{\nu}\}$ are also \emph{stochastically} independent  uniformly distributed random variables over $\GFqp$ ($\tilde e_i$ is uniformly distributed over $\GFqp$ for all $i$ and $\tilde{\bm e}$ is uniformly distributed over $\GFqp^{\nu}$). Finally, since $e \in \{e_1,\ldots,e_w\} \setminus \{\tilde e_1,\dots,\tilde e_{\nu}\}$ can be written as a $\GFq$-linear combination of $\{\tilde e_1,\dots,\tilde e_{\nu}\}$, it follows that $H(\bm e) = H(\tilde{\bm e})= \nu \cdot p \log q$.
From Lemma~\ref{Lemma: entropy}, $H(\bm e) \leq H(\bm r)$. Thus, $u \geq \nu$ since 
$H(\bm e) = \nu \cdot p \log q$ and $H(\bm r) = u \cdot p \log q$. 
The conditional entropy $H(\bm r|\bm e,\bm m)$ is equal to the logarithm (base-$2$) of the number of solutions of (\ref{eq:fy}) when the number of unknowns $u$ is larger than or equal to the number of independent equations $\nu$, i.e.,  when $u \geq \nu$. Hence, 
		$H(\bm r|\bm e,\bm m)=(u-\nu)p\log q = H(\bm r)-H(\bm e)$
from which the result follows from (\ref{Eq: lemma}) (see Theorem~\ref{Lemma: Secrecy lemma}). 
\end{IEEEproof}


\begin{example}
Consider the $(20,10)$ secure RFC over $GF(q^p)$ in Fig.~\ref{Fig: Secure repairable fountain code} that encodes the message $\bm m=(m_1,\ldots,m_6)$ of $6$ symbols and a vector $\rr=(r_1,\ldots,r_4)$ of $4$ random symbols. Each $c_i, i=1,\ldots, 20$, is an evaluation of a linearized polynomial $f(\cdot)$ at $y_i$. 
For the $(1,1)$ eavesdropper model, the scenario where $\S_1=\{c_6\}$ and $\S_2=\{c_5\}$, i.e., the eavesdropper gains access to the symbols $\bm e=(c_5,c_6,c_8, c_{18}= c_5+ c_6+ c_8)$, is depicted. It can easily be seen that $H(\bm r)=4p\log q$, $H(\bm e)=3p\log q$, and  $H(\bm r|\bm e,\bm m)=(4-3)p\log q$. Therefore, there is no information leakage to the eavesdropper.
%
\end{example}

\section{Numerical Results}
\begin{figure}
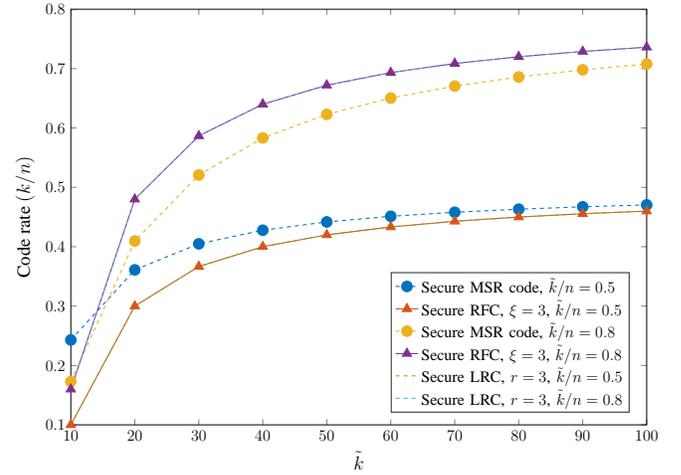

	\centering
	\include{Plot_coderate}
\vspace{-3ex}
	\caption{Comparison of code rates for different classes of secure ECCs for the $(2,2)$ eavesdropper model.}
	\label{Fig: plot}
	\vspace{-2ex}
\end{figure}

We compare the proposed secure RFCs with the secure MSR codes and secure LRCs in \cite{Raw14} in terms of the maximum code rate $k/n$ that allows to achieve security. In particular, we consider $(r,\delta)$ $d_{\text{min}}-$optimal LRCs \cite{Raw14}, where $r$ is the code locality (and thus has an analogous meaning to $\xi$ for RFCs) and $d_{\text{min}}$ is the minimum distance of the code. Each local group of such a code can be seen as a subcode (punctured from the LRC) of minimum distance at least $\delta$.




In Fig.~\ref{Fig: plot}, we fix the code rate of the inner code (RFC, LRC, or MSR code), $\ktilde/n$, to $0.5$ and $0.8$, and plot the achievable code rates (the maximum $k/n$ that allows to achieve security) for the $(2,2)$ eavesdropper model as a function of $\ktilde$. Note that $\ktilde/n$ is an upper bound on the achievable code rate $k/n$, since to achieve security a number of random symbols needs to be appended to the message of length $k$. Note also that $n$ is the total number of storage nodes. We remark that, unlike LRC- and RFC-based DSSs, where each code symbol is stored in a different storage node, for MSR codes each storage node stores $\alpha=(n-\tilde{k})^{\tilde k-1}$ code symbols. For a fair comparison between RFCs and LRCs, we set $r=\xi$ and $\delta=2$. It can be seen that the achievable code rates for secure RFCs and secure LRCs are identical. On the other hand, secure RFCs yield higher achievable code rates compared to secure MSR codes for $\ktilde/n=0.8$, while the opposite is observed for $\ktilde/n=0.5$.

\section{Conclusion}
We proposed a code construction based on RFCs that is secure against the $(\ell_1,\ell_2)$ eavesdropper model. We gave a necessary and sufficient condition for the information leakage to the eavesdropper to be zero, and subsequently proved that the proposed construction is completely secure. The proposed secure RFCs yield the same achievable code rates as LRCs, and higher than MSR codes (when the code rate of the underlying code is high enough). An interesting extension of this work would be the design of secure and repair-efficient vector RFCs, i.e., code symbols are distributed over the storage nodes, each containing $\alpha>1$ code symbols.




%

%
%
%
%
%

\ifCLASSOPTIONcaptionsoff
  \newpage
\fi



\bibliographystyle{IEEEtran}
%

\balance
\bibliography{Bib_SRFC}

%

%






\end{document}

%% file: Example_Introduction.tex
\begin{tikzpicture}[thick, scale=0.5, every node/.style={transform shape}]
	\Large
	\begin{scope}[shift={(0,-2)}]
		\node[draw, rectangle, minimum size=1cm] (c1) at (0.5,8.5) {$c_1$};
		\node[draw, rectangle, minimum size=1cm] (c2) at (2,8.5) {$c_2$};
		\node[draw, rectangle, minimum size=1cm] (c3) at (3.5,8.5) {$c_3$};
		\node[draw, rectangle, minimum size=1cm] (c4) at (5,8.5) {$c_4$};
	\end{scope}
	
	\node[draw, circle, minimum size=1cm, black] (m1) at (0.5,4.5) {$m_1$};
	\node[draw, circle, minimum size=1cm, black] (m2) at (2,4.5) {$m_2$};
	\node[draw, circle, minimum size=1cm] (m3) at (3.5,4.5) {$m_3$};
	\node[draw, circle, minimum size=1cm, black] (m4) at (5,4.5) {$m_4$};
	
	\begin{scope}[shift={(1.5,2)}]
		\node[draw, rectangle, minimum size=1cm, black] (c5) at (0.5,0.5) {$c_{5}$};
		\node[draw, rectangle, minimum size=1cm] (c6) at (2,0.5) {$c_{6}$};
	\end{scope}
	
	\draw[black] (m1)--(c1);
	\draw (m2)--(c2);
	\draw (m3)--(c3);
	\draw (m4)--(c4);
	
	\draw (m1)--(c6.north);
	\draw[black] (m2)--(c5);
	\draw (m3)--(c6);
	\draw[black] (m4)--(c5.north);
\end{tikzpicture}

%% file: Example_EavesdropperModel.tex
\begin{tikzpicture}[thick, scale=0.5, every node/.style={transform shape}, gray]
	\Large
	\begin{scope}[shift={(0,-2)}]
		\node[draw, rectangle, minimum size=1cm, black,fill=red!50] (c1) at (0.5,8.5) {$c_1$};
		\node[draw, rectangle, minimum size=1cm, black,fill=red!50] (c2) at (2,8.5) {$c_2$};
		\node[draw, rectangle, minimum size=1cm] (c3) at (3.5,8.5) {$c_3$};
		\node[draw, rectangle, minimum size=1cm, black] (c4) at (5,8.5) {$c_4$};
	\end{scope}
	
	\node[draw, circle, minimum size=1cm, black, fill=gray!50] (m1) at (0.5,4.5) {$m_1$};
	\node[draw, circle, minimum size=1cm, black, fill=gray!50] (m2) at (2,4.5) {$m_2$};
	\node[draw, circle, minimum size=1cm] (m3) at (3.5,4.5) {$m_3$};
	\node[draw, circle, minimum size=1cm, black, fill=gray!50] (m4) at (5,4.5) {$m_4$};
	
	\begin{scope}[shift={(1.5,2)}]
		\node[draw, rectangle, minimum size=1cm, black, fill=gray!50] (c5) at (0.5,0.5) {$c_{5}$};
		\node[draw, rectangle, minimum size=1cm] (c6) at (2,0.5) {$c_{6}$};
	\end{scope}
	
	\draw[black] (m1)--(c1);
	\draw[black] (m2)--(c2);
	\draw (m3)--(c3);
	\draw[black] (m4)--(c4);
	
	\draw (m1)--(c6.north);
	\draw[black] (m2)--(c5);
	\draw (m3)--(c6);
	\draw[black] (m4)--(c5.north);
\end{tikzpicture}

%% file: Example_securityV_2.tex
\newlength{\hatchspread}
\newlength{\hatchthickness}
\newlength{\hatchshift}
\newcommand{\hatchcolor}{}
	\begin{tikzpicture}[thick, scale=0.5, every node/.style={transform shape}, gray, inner sep=1]
	\Large
	\tikzset{hatchspread/.code={\setlength{\hatchspread}{#1}},
	         hatchthickness/.code={\setlength{\hatchthickness}{#1}},
	         hatchshift/.code={\setlength{\hatchshift}{#1}},
	         hatchcolor/.code={\renewcommand{\hatchcolor}{#1}}}
	
	\tikzset{hatchspread=3pt,
	         hatchthickness=0.4pt,
	         hatchshift=0pt,
	         hatchcolor=black}
	
	\pgfdeclarepatternformonly[\hatchspread,\hatchthickness,\hatchshift,\hatchcolor]
	   {custom north west lines}
	   {\pgfqpoint{\dimexpr-2\hatchthickness}{\dimexpr-2\hatchthickness}}
	   {\pgfqpoint{\dimexpr\hatchspread+2\hatchthickness}{\dimexpr\hatchspread+2\hatchthickness}}
	   {\pgfqpoint{\dimexpr\hatchspread}{\dimexpr\hatchspread}}
	   {
	    \pgfsetlinewidth{\hatchthickness}
	    \pgfpathmoveto{\pgfqpoint{0pt}{\dimexpr\hatchspread+\hatchshift}}
	    \pgfpathlineto{\pgfqpoint{\dimexpr\hatchspread+0.15pt+\hatchshift}{-0.15pt}}
	    \ifdim \hatchshift > 0pt
	      \pgfpathmoveto{\pgfqpoint{0pt}{\hatchshift}}
	      \pgfpathlineto{\pgfqpoint{\dimexpr0.15pt+\hatchshift}{-0.15pt}}
	    \fi
	    \pgfsetstrokecolor{\hatchcolor}
	    \pgfusepath{stroke}}

		\begin{scope}[shift={(0,-2)}]
		\node[draw, rectangle, minimum size=1cm] (c1) at (0.5,8.5) {$c_1$};
		\node[draw, rectangle, minimum size=1cm] (c2) at (2,8.5) {$c_2$};
		\node[draw, rectangle, minimum size=1cm] (c3) at (3.5,8.5) {$c_3$};
		\node[draw, rectangle, minimum size=1cm] (c4) at (5,8.5) {$c_4$};
		\node[draw, rectangle, minimum size=1cm, black, fill=red!50] (c5) at (6.5,8.5) {$c_5$};
		\node[draw, rectangle, minimum size=1cm, black, fill=red!50] (c6) at (8,8.5) {$c_6$};
		\node[draw, rectangle, minimum size=1cm] (c7) at (9.5,8.5) {$c_7$};
		\node[draw, rectangle, minimum size=1cm, black] (c8) at (11,8.5) {$c_8$};
		\node[draw, rectangle, minimum size=1cm] (c9) at (12.5,8.5) {$c_9$};
		\node[draw, rectangle, minimum size=1cm] (c10) at (14,8.5) {$c_{10}$};
		\end{scope}
		
		\node[draw, rectangle, minimum size=1cm] (c11) at (0.5,0.5) {$c_{11}$};
		\node[draw, rectangle, minimum size=1cm] (c12) at (2,0.5) {$c_{12}$};
		\node[draw, rectangle, minimum size=1cm] (c13) at (3.5,0.5) {$c_{13}$};
		\node[draw, rectangle, minimum size=1cm] (c14) at (5,0.5) {$c_{14}$};
		\node[draw, rectangle, minimum size=1cm] (c15) at (6.5,0.5) {$c_{15}$};
		\node[draw, rectangle, minimum size=1cm] (c16) at (8,0.5) {$c_{16}$};
		\node[draw, rectangle, minimum size=1cm] (c17) at (9.5,0.5) {$c_{17}$};
		\node[draw, rectangle, minimum size=1cm, black, fill=gray!30] (c18) at (11,0.5) {$c_{18}$};
		\node[draw, rectangle, minimum size=1cm] (c19) at (12.5,0.5) {$c_{19}$};
		\node[draw, rectangle, minimum size=1cm] (c20) at (14,0.5) {$c_{20}$};
		
		\node[draw, circle, minimum size=1cm] (f1) at (0.5,4.5) {$\tilde c_1$};
		\node[draw, circle, minimum size=1cm] (f2) at (2,4.5) {$\tilde c_2$};
		\node[draw, circle, minimum size=1cm] (f3) at (3.5,4.5) {$\tilde c_3$};
		\node[draw, circle, minimum size=1cm] (f4) at (5,4.5) {$\tilde c_4$};
		\node[draw, circle, minimum size=1cm, black,  fill=gray!30] (f5) at (6.5,4.5) {$\tilde c_5$};
		\node[draw, circle, minimum size=1cm, black,  fill=gray!30] (f6) at (8,4.5) {$\tilde c_6$};
		\node[draw, circle, minimum size=1cm] (f7) at (9.5,4.5) {$\tilde c_7$};
		\node[draw, circle, minimum size=1cm, black,  fill=gray!30] (f8) at (11,4.5) {$\tilde c_8$};
		\node[draw, circle, minimum size=1cm] (f9) at (12.5,4.5) {$\tilde c_9$};
		\node[draw, circle, minimum size=1cm] (f10) at (14,4.5) {$\tilde c_{10}$};
		
		\draw (f1)--(c1);
		\draw (f2)--(c2);
		\draw (f3)--(c3);
		\draw (f4)--(c4);
		\draw[black] (f5)--(c5);
		\draw[black] (f6)--(c6);
		\draw (f7)--(c7);
		\draw[black] (f8)--(c8);
		\draw (f9)--(c9);
		\draw (f10)--(c10);
		
		\draw (c11.north)--(f5.south);
		\draw (c11.north)--(f6.south);
		\draw (c11.north)--(f8.south);
		
		\draw (c12.north)--(f2.south);
		\draw (c12.north)--(f4.south);
		\draw (c12.north)--(f7.south);
		
		\draw (c13.north)--(f2.south);
		\draw (c13.north)--(f3.south);
		\draw (c13.north)--(f4.south);
		
		\draw (c14.north)--(f1.south);
		\draw (c14.north)--(f3.south);
		\draw (c14.north)--(f4.south);
		
		\draw (c15.north)--(f2.south);
		\draw (c15.north)--(f5.south);
		\draw (c15.north)--(f6.south);
		
		\draw (c16.north)--(f2.south);
		\draw (c16.north)--(f6.south);
		\draw (c16.north)--(f9.south);
		
		\draw (c17.north)--(f4.south);
		\draw (c17.north)--(f6.south);
		\draw (c17.north)--(f10.south);
		
		\draw[black] (c18.north)--(f5.south);
		\draw[black] (c18.north)--(f6.south);
		\draw[black] (c18.north)--(f8.south);
		
		\draw (c19.north)--(f1.south);
		\draw (c19.north)--(f2.south);
		\draw (c19.north)--(f6.south);
		
		\draw (c20.north)--(f1.south);
		\draw (c20.north)--(f2.south);
		\draw (c20.north)--(f10.south);
		
		\begin{scope}[shift={(0,-2)}]
		
		\end{scope}
	\end{tikzpicture}

%% file: Plot_coderate.tex
\definecolor{mycolor1}{rgb}{0.00000,0.44700,0.74100}%
\definecolor{mycolor2}{rgb}{0.85000,0.32500,0.09800}%
\definecolor{mycolor3}{rgb}{0.92900,0.69400,0.12500}%
\definecolor{mycolor4}{rgb}{0.49400,0.18400,0.55600}%
\definecolor{mycolor5}{rgb}{0.46600,0.67400,0.18800}%
\definecolor{mycolor6}{rgb}{0.30100,0.74500,0.93300}%
\begin{tikzpicture}[thick, scale=0.5, every node/.style={transform shape}]
\large
\begin{axis}[%
width=6.028in,
height=4.354in,
at={(1.011in,0.642in)},
scale only axis,
xmin=10,
xmax=100,
xlabel={\Large $\tilde k$},
ymin=0.1,
ymax=0.8,
ylabel={\Large Code rate $(k/n)$},
axis background/.style={fill=white},
legend style={at={(0.97,0.03)},anchor=south east,legend cell align=left,align=left,draw=white!15!black},
mark size=4pt,
]
\addplot [color=mycolor1,dashed,mark=*,mark options={solid},thick]
  table[row sep=crcr]{%
10	0.243\\
20	0.361\\
30	0.404925925925926\\
40	0.42778125\\
50	0.441784\\
60	0.451240740740741\\
70	0.458055393586006\\
80	0.46319921875\\
90	0.467219478737997\\
100	0.470448\\
};
\addlegendentry{Secure MSR code, $\tilde k/n =0.5$};

\addplot [color=mycolor2,solid,mark=triangle*,mark options={solid},thick]
  table[row sep=crcr]{%
10	0.1\\
20	0.3\\
30	0.366666666666667\\
40	0.4\\
50	0.42\\
60	0.433333333333333\\
70	0.442857142857143\\
80	0.45\\
90	0.455555555555556\\
100	0.46\\
};
\addlegendentry{Secure RFC, $\xi=3$, $\tilde k/n=0.5$};

\addplot [color=mycolor3,dashed,mark=*,mark options={solid},thick]
  table[row sep=crcr]{%
10	0.1728\\
20	0.4096\\
30	0.52077037037037\\
40	0.5832\\
50	0.6229504\\
60	0.65042962962963\\
70	0.670544606413994\\
80	0.6859\\
90	0.698003840877915\\
100	0.7077888\\
};
\addlegendentry{Secure MSR code, $\tilde k/n=0.8$};

\addplot [color=mycolor4,solid,mark=triangle*,mark options={solid},thick]
  table[row sep=crcr]{%
10	0.16\\
20	0.48\\
30	0.586666666666667\\
40	0.64\\
50	0.672\\
60	0.693333333333333\\
70	0.708571428571429\\
80	0.72\\
90	0.728888888888889\\
100	0.736\\
};
\addlegendentry{Secure RFC, $\xi=3$, $\tilde k/n=0.8$};

\addplot [color=mycolor5,dashed,thick]
  table[row sep=crcr]{%
10	0.1\\
20	0.3\\
30	0.366666666666667\\
40	0.4\\
50	0.42\\
60	0.433333333333333\\
70	0.442857142857143\\
80	0.45\\
90	0.455555555555556\\
100	0.46\\
};
\addlegendentry{Secure LRC, $r=3$, $\tilde k/n=0.5$};

\addplot [color=mycolor6,dashed,thick]
  table[row sep=crcr]{%
10	0.16\\
20	0.48\\
30	0.586666666666667\\
40	0.64\\
50	0.672\\
60	0.693333333333333\\
70	0.708571428571429\\
80	0.72\\
90	0.728888888888889\\
100	0.736\\
};
\addlegendentry{Secure LRC, $r=3$, $\tilde k/n=0.8$};

\end{axis}
\end{tikzpicture}%